\documentclass{article}
\usepackage[dvips]{graphicx}
\usepackage{epsfig}

\begin{document}

\title{Paradoxical games and Brownian thermal engines}
\author{Juan M.R. Parrondo$^{1}$ and Borja Jimenez de Cisneros$^{1}$\\
$^1$ Dep. F\'{i}sica At\'omica, Molecular y Nuclear \\
Universidad
Complutense de Madrid 28040 Madrid, Spain\\
\\
Translated from Spanish into English by:\\
Pau Amengual (\S), Juan Sotelo (\dag) and Derek Abbott (\ddag)\\
\\
(\S)\small{ Instituto Mediterr\'aneo de Estudios Avanzados, IMEDEA
(CSIC-UIB)}\\\small{Ed. Mateau Orfila, Campus UIB, E-07122 Palma
de Mallorca, Spain}\\
\\
(\dag) \small{Dept of Materials Science, The \AA ngstr\"{o}m
Laboratory}\\\small{Uppsala University, P.O. Box 534, SE-751 21 Uppsala, Sweden}\\
\\
(\ddag)\small{Centre for Biomedical Engineering (CBME) and}\\
\small{Department of Electrical and Electronic Eng.}\\
\small{The University of Adelaide, SA 5005, Australia}}
\date{}
\maketitle

\begin{abstract}{}
Two losing games, when alternated in a periodic or random fashion,
can produce a winning game. This paradox occurs in a family of
stochastic processes: if one combines two or more dynamics where a
given quantity decreases, the result can be a dynamic system where
this quantity increases. The paradox could be applied to a number
of stochastic systems and has drawn the attention of researchers
from different areas. In this paper we show how the phenomenon can
be used to design Brownian or molecular motors, i.e., thermal
engines that operate by rectifying fluctuations. We briefly review
the literature on Brownian motors, pointing out that a new
thermodynamics of Brownian motors will be fundamental to the
understanding of most processes of energy transduction in
molecular biology.
\end{abstract}

\section{Paradoxical games}

Suppose we have a biased coin, so that the probability that any
flip will result in a head is $0.5-\epsilon$, where $\epsilon$ is
a small and positive number. With this coin, call it coin 1, it is
proposed that you play the following game: if the flip results in
a tail you receive $1$ dollar, if not you pay the same amount. If
you have unlimited capital then you should accept to play the game
without hesitation: if $x(t)$ is your opponent's capital after $t$
runs, it will not be difficult for you to show that its mean
value, $\langle{x(t)}\rangle$ is a strictly a decreasing function
of $t$ (while your average capital is an increasing function of
$t$).

\begin{figure}[!htb]
\centerline{\epsfig{figure=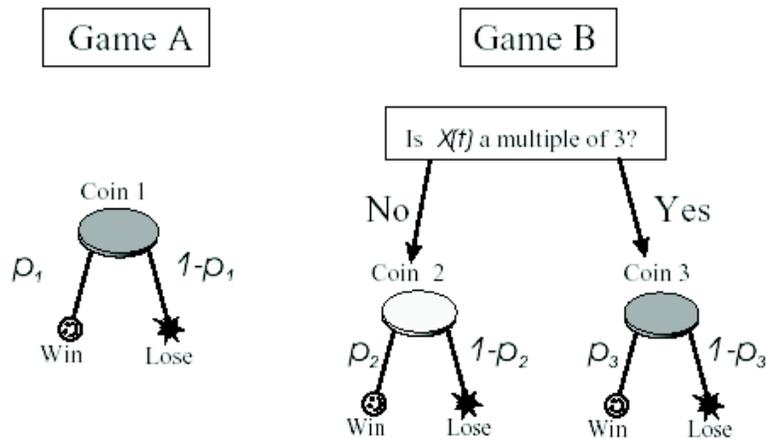,width=10.5cm}}
 \caption{Rules of games A and B. The values
of the probabilities are: $p_1 =\frac{1}{2}-\epsilon$,
$p_2=\frac{3}{4}-\epsilon$ and $p_3=\frac{1}{10}-\epsilon$, with
$\epsilon$ small and positive, which in all cases favour the
``losing" option. The lighter colour of coin $2$ indicates that
with this coin the probability of winning is greater than
$\frac{1}{2}$ and so it is a favourable coin.} \label{coins}
\end{figure}

Now, it is proposed that you play a second game that uses two
coins, call them coin $2$ and coin $3$ respectively. Coin $3$ is
flipped whenever your opponent's capital is a multiple of $3$, in
any other case coin $2$ is flipped (note that your opponent can
have negative capital; by a multiple of $3$ it is understood any
integer number that can be written as $3n$, with $n$ an integer
number). Let $p_2=0.75-\epsilon$ and $p_3=0.1-\epsilon$ be the
probabilities that your opponent wins with coin $2$ and coin $3$
respectively, see Figure \ref{coins}. The analysis of this game is
not as simple as that of our previous game. However, it can be
shown -- see Appendix 1 and Figure \ref{coins} -- that this game,
too, is favourable for you in the sense that the mean value of the
opponent's capital $\langle{x(t)}\rangle$ is, once again, a
strictly decreasing function of $t$. Let us call the first game we
have described ``game A'' and the second one ``game B.''

Once you are convinced that your opponent will lose in both games
you are given a third proposal: alternate the games following the
sequence AABBAABB... If you frown, the proposal can be modified to
make it less suspicious: in each run we will randomly chose the
game that is played. If you accept either of these proposals you
would have trusted your intuition too much, not realizing that
random systems, even as simple as the ones we have just described,
may behave in an unexpected way.

\begin{figure}[!htb]
\centerline{\epsfig{figure=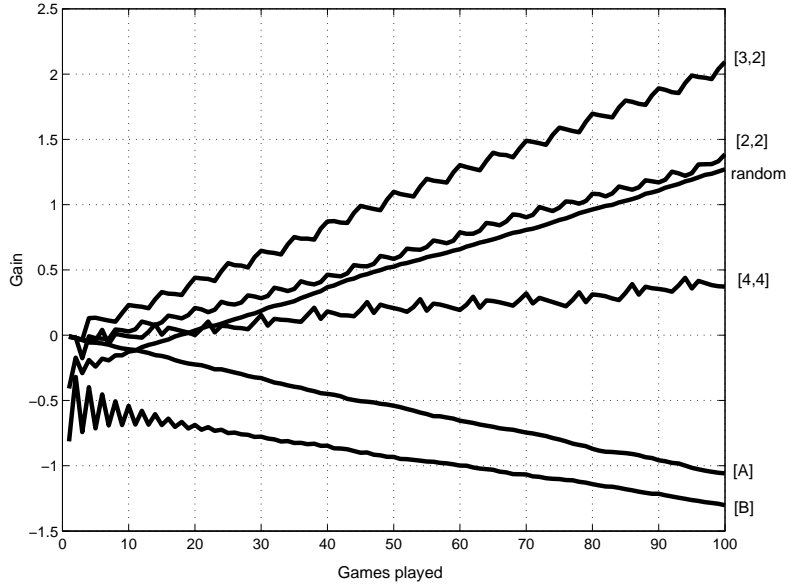,width=10.5cm}}
\caption{Average capital (over $50\,000$ players), as a function
of the number of runs, for each of the games and in several
combinations, including a random combination. In all cases
$\epsilon=0.005$ and we have used the notation $[a,b]$ to indicate
that we play $a$ runs of game A, followed by $b$ runs of game B,
and so on. The random combination does not differ much from the
combination $[2,2]$, that is, the combinaton AABBAA...(After
\cite{harmer99a}).} \label{game_plot}
\end{figure}

Indeed, if we alternate the games, either by following a fixed
sequence AABBAABB... or chosing randomly the game we play, then
your opponent's average capital $\langle{x(t)}\rangle$ will be a
strictly increasing function of $t$. Figure \ref{game_plot} shows
your opponent's capital in different situations: games A and B
played alone, in several periodic combinations and in a random
combination.

\section{Detailed analysis of the games}

The phenomenon we have just described is known as Parrondo's
paradox and it is receiving some attention
\cite{harmer99a,harmer99b} because it is thought it might have
application in areas such as economics and evolutionary theory. It
is true that this phenomenon may arise in any situation that
involves the interplay of two or more random dynamics. However, we
have not yet described a real situation where the paradox takes
place.

Let us see how we can analyse these paradoxical games. Game B and
the random combination of games A and B may be reduced to a
3-state Markov chain \cite{karlin75}. These states are: the
capital is multiple of $3$, multiple of $3$ plus $1$ or multiple
of $3$ plus $2$. The variable that determines these states is

\begin{equation}
\label{equation1} y\equiv x \hspace{0.3cm} \mathrm{mod}\,3
\end{equation}

which only takes on the values $0$, $1$ and $2$. The analysis of
this Markov chain can be carried out by diagonalizing a
$3{\times}3$ matrix (see Appendix 1).

This analysis gives us the following intuitive explanation of the
paradox. Game B uses two coins: one ``bad", coin $3$, and one
``good", coin $2$. When game B is played alone, the probability of
using coin $3$ is

\begin{equation}
\label{equation2}
\pi_0=\frac{5}{13}-\frac{440}{2197}\epsilon+\ldots
\end{equation}

where we have dropped the terms of order $\epsilon^2$ to simplify
the analysis. Notice, also, that the probability $\pi_0$ is not
equal to $\frac{1}{3}$, as one might have initially thought, but
greater than $\frac{1}{3}$. Consequently, the probability of
winning is

\begin{equation}
\label{equation3}
P_{\mathrm{win}}=\pi_0p_3+(1-\pi_0)p_2=\frac{1}{2}-\frac{147}{169}\epsilon+\ldots
\end{equation}

which is less than $\frac{1}{2}$ for any positive $\epsilon$.
Nevertheless, when games A and B are combined randomly, the
probability of using coin $3$ becomes

\begin{equation}
\label{equation4}
\pi'_0=\frac{245}{709}-\frac{48880}{502681}\epsilon+\ldots
\end{equation}

which is less than $\pi_0$. The probability of winning is now

\begin{equation}
\label{equation5}
P'_{\mathrm{win}}=\pi'_0\frac{p_3+p_1}{2}+(1-\pi'_0)\frac{p_2+p_1}{2}=
\frac{727}{1418}-\frac{486795}{502681}\epsilon+\ldots
\end{equation}

which is greater than $\frac{1}{2}$ if $\epsilon$ is small enough.
What happens then is that game A, even though it consists of just
one ``bad" coin, redistributes the frequencies of use of the two
coins of game B in such a way that the ``good" coin is being used
more often than its counterpart. This is the essence of the
paradox: the winning tendency is already built-in into game B, but
when this game is played alone, the losing tendency is dominant.
The role of game A is to reverse this dominancy. Thus, game A, in
spite of being a losing game, effects a reprise of the ``good"
coin of game B that overcomes its own losing tendency, so that the
combination of games A and B becomes a winning game.

\section{A simple Brownian motor}

What has all this to do with physics? Even though now the paradox
is part of probability theory, it was originally inspired by a
physical system, namely, a Brownian motor or molecular motor
\cite{astumian94,sokolov97}.

With a slight change in the winning and losing probabilities, the
games described above could be interpreted as a Brownian particle
in one-dimension subjected to certain potentials.

\begin{figure}[!htb]
\centerline{\epsfig{figure=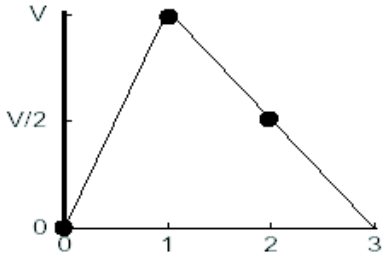,width=5.5cm}} \caption{The
potential used on the Brownian particle described in this
article.} \label{potential}
\end{figure}

Let us consider the following model of a Brownian particle in
one-dimension,  where the particle can occupy the discrete set of
positions $x=0, \pm 1 \pm 2,\ldots$.  The particle jumps with
certain probability from one position to another at discrete times
$t=0, 1, 2, ...$ Suppose the particle is under the action of a
certain periodic potential $V(x)$ of period $3$, i.e.
$V(x+3)=V(x)$. This means the potential is completely defined by
the three values: $V(0)$, $V(1)$ and $V(2)$, which we denote by
$V_0$, $V_1$ and $V_2$, respectively. We will set $V_0=0$, $V_1=V$
and $V_2=\frac{V}{2}$, so that the potential is asymmetric (see
Figure \ref{potential}). In addition, suppose there is an external
force $F$ acting on the particle and pointing to the left. The
energy at each point is then

\begin{equation}
\label{equation6} E(x)=V(x)+Fx.
\end{equation}

How does this particle move at a certain temperature $T$? When a
physical system is at a certain temperature $T$ and can be found
in different states $i$ with energy $E_i$ its evolution may be
described by a Markov chain \cite{koonin75}. The dynamics are
expressed in terms of the transition probabilities $p(i\rightarrow
j)$ for jumping from state $i$ to the state $j$. These
probabilities must satisfy the detailed-balance condition. One of
the dynamics to satisfy this requirement is the Metropolis
algorithm (see Appendix 2) that is used in Monte-Carlo simulations
of systems in equilibrium at a certain temperature $T$
\cite{koonin75}.

\begin{figure}[!htb]
\centerline{\epsfig{figure=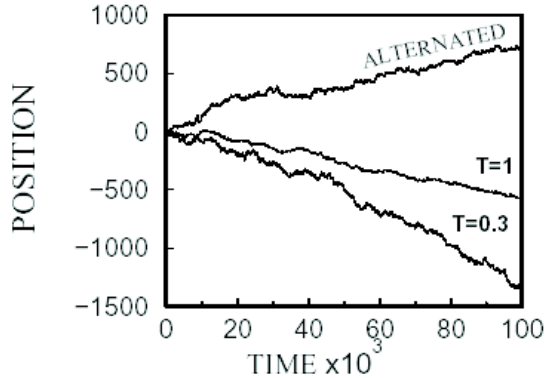,width=7.5cm}} \caption{The
position of the Brownian particle for the two temperatures,
$T_1=1$ and $T_2=0.3$ and the alternating sequence
$T_1T_1T_2T_2T_1T_1$. Just like in the games (Figure
\ref{game_plot}) we clearly see the inversion of [the direction
of] movement when we alternate the two temperatures.}
\label{position}
\end{figure}

When our Brownian particle evolves following the Metropolis
algorithm at a finite temperature $T$, its position $x(t)$ behaves
more or less like the capital in game B. On the other hand, if we
follow the Metropolis algorithm at a very high temperature, the
resulting system is very much like game A. In both cases the
external force plays the role of the games' $\epsilon$ parameter.
These analogies give us a strong indication how to reproduce the
paradox in the behaviour of the Brownian particle. At any
temperature the particle moves to the left because this is the
direction of the external force $F$. Hence, the mean value of
$\langle{x(t)}\rangle$ decreases with $t$. On the other hand, if
we alternate two different temperatures $T_1$ and $T_2$, the
particle moves to the right against the external force, that is
$\langle{x(t)}\rangle$ increases with $t$. Again, two dynamics
where $\langle{x(t)}\rangle$ decreases give rise, when combined,
to dynamics where $\langle{x(t)}\rangle$ increases. This
phenomenon is depicted in Figure \ref{position}; there we plot the
position of the particle as a function of  time, obtained by
numerical simulation for the temperatures, $T_1=1$ and $T_2=0.3$,
and the alternating sequence $T_1T_1T_2T_2T_1T_1$.

Even more interesting is to check that the particle is a thermal
engine. Let us consider a periodic alternating sequence of the
type $T_1T_1T_2T_2T_1T_1\ldots$, that is, we put the particle in
contact with a thermal bath at temperature $T_1$ for two
time-steps, then in contact with a thermal bath at temperature
$T_2$ for the next two time-steps, and so on. Let us assume
$T_1>T_2$. We have, thus, a system in contact with two thermal
baths at different temperatures just as in the well known cycles
of the classical thermal engines. So, the particle behaves exactly
as a thermal engine, that is, it extracts energy from the hot
source, does work against the external force and dissipates part
of the extracted energy in the form of heat towards the cold
source.

Let us see how to  quantitatively study the energy exchange of
this engine. In a certain step $t\rightarrow t+1$, in which the
particle is in contact with, say, bath $1$, the energy transferred
from the bath to the particle is

\begin{equation}
\label{equation7}
Q^{(1)}_{t\rightarrow t+1}=
\sum^{\infty}_{x=-\infty}[\pi_{x}(t+1)-\pi_{x}(t)](V(x)+Fx)
\end{equation}

where $\pi_x(t)$ is the probability that the particle is at the
position $x$ in time $t$. The right hand side of the above
equation can be decomposed into two terms

\begin{equation}
\label{equation8} Q^{(1)}_{t\rightarrow
t+1}=\triangle\mathcal{U}_{t\rightarrow t+1}-W_{t\rightarrow t+1}.
\end{equation}

The first term is $\triangle\mathcal{U}_{t\rightarrow
t+1}=\mathcal{U}_{t+1}-\mathcal{U}_{t}$, where $\mathcal{U}(t)$ is
the internal energy of the system

\begin{equation}
\label{equation9}
 \mathcal{U}(t)=\sum^{\infty}_{x=-\infty} \pi_{x}(t)V(x)=\langle V(x(t))
 \rangle.
\end{equation}

On the other hand, the second term is the change of energy due to
the external force $F$, that is, the work done by the force on the
particle

\begin{equation}
\label{equation10} W_{t\rightarrow t+1}=-F\langle v(t) \rangle
\end{equation}

where $\langle{v}\rangle = \langle{x(t+1)}\rangle -
\langle{x(t)}\rangle$ is the mean velocity of the particle. Hence
Equation (\ref{equation10}) has a simple interpretation: the work
done by the force on the particle in a unit of time, or the power
developed by the force, is the product of the force times the mean
velocity of the particle. The minus sign arises because the
direction of the force points to the left.

\begin{figure}[!htb]
\centerline{\epsfig{figure=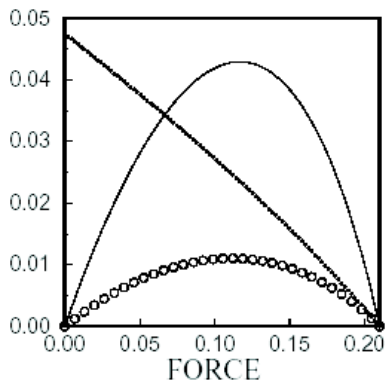,width=5.5cm}} \caption{The
efficiency $\eta$ of the thermal engine (continuous line), work
$W$ done in each cycle (circles) and the current of particles
(dotted line) as a function of the external force $F$ for $T_1=1$,
$T_2=0.1$ and $V=1$.} \label{force}
\end{figure}

The three quantities we have defined, the internal energy, heat
and work are related through Equation (\ref{equation8}), which is
nothing else that the First Law of Thermodynamics\footnote{An
alternative viewpoint considers the term $-Fx(t)$ as part of the
internal energy. In this interpretation, the heat dissipated in
each bath is the same as the one computed in this article. There
are changes, however, in the work, which is always zero, and the
internal energy. This interpretation has two problems. Firstly,
the engine is not strictly cyclic, because the internal energy
increases in each cycle. And secondly, it is not possible to
define in a simple way the efficiency of the engine.} (the sign
convention we follow is the usual one in physics, namely, that
heat and work are both positive if there is energy entering the
system).

The same procedure is applied to the time-steps when the system is
in contact with bath $2$. When we alternate the two baths in a
cycle of four steps, $T_1T_1T_2T_2\ldots$, the system, after a
certain number of cycles, reaches a regime where the internal
energy and the mean velocity are periodic in time. In this regime
we can compute the amount of heat extracted from each bath and the
total work done by the force. The analytical treatment is
analogous to the one carried out for the games and it is based on
a three-state Markov chain. We shall not include here the details
of the calculations but shall give some of the results that can be
obtained. Applying equation (\ref{equation7}) to the first two
steps of the cycle we compute the energy $Q_1$ transferred from
bath $1$ to the system, which is positive if $T_1>T_2$. However,
when we apply the same equation to the last two steps of the cycle
we obtain a negative energy $Q_2$, and, if $F$ is weak enough, we
obtain a negative $W= -Q_1-Q_2$. These results tell us that the
system extracts energy from the hot bath, dissipates part of this
energy towards the cold bath and uses what remains of it to do
work against the external force. The efficiency of the thermal
engine is

\begin{equation}
\label{equation11}
\eta=-\frac{W}{Q_1}=1+\frac{Q_2}{Q_1}
\end{equation}

which is positive and less than 1.

In Figure \ref{force} we plot the engine efficiency as a function
of the external force $F$. When the force is zero the efficiency
is zero because the engine does not do any work and there is an
irreversible heat transfer from the hot source to the cold one.
Note that this holds even though in this case there exists a
current of particles. For sufficiently strong forces the
efficiency is once again zero. The reason is that the particles
cannot move against strong forces. On the other hand, there exists
a force for which the particle is, on the average, at rest and
does not do work, while there still exists heat transfer from the
hot bath to the cold one.

\begin{figure}[!htb]
\centerline{\epsfig{figure=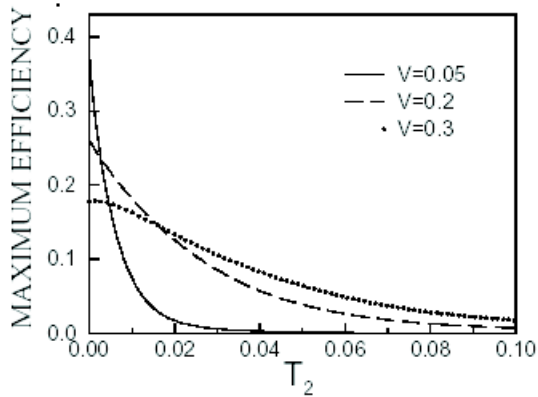,width=7.5cm}}
 \caption{Efficiency of the thermal
engine as a funtion of the temperature of the cold source $T_2$,
for $T_1=1$ and several values of $V$. For each point, we have
taken the value of the external force that maximizes the
efficiency.} \label{efficiency}
\end{figure}

As it can be seen, the engine always works irreversibly, so it can
never attain the efficiency of the Carnot engine. This is a common
fact in the majority of Brownian motors
\cite{sokolov97,parrondo98b} (an exception are the reversible
ratchets \cite{parrondo98a}, whose efficiencies can be arbitrarily
close to Carnot's \cite{parrondo98b}), including the Feynman
ratchet \cite{feynman63}. It is worth pointing out here that
Feynman did not notice the intrinsic irreversibility of his
thermal engine and went on to show that under a certain hypothesis
that the engine could attain the Carnot efficiency. It has
recently been shown that such hypotheses are incorrect
\cite{parrondo96}.

In Figure \ref{efficiency} we show the engine's efficiency for
several temperatures $T_2$ keeping fixed the temperature $T_1$.
Observe that the Carnot efficiency curve,
$1-\frac{T_2}{T_1}=1-T_2$, is well above all the curves plotted in
Figure \ref{efficiency}.

\section{Conclusions}

We have shown a physical application of the paradoxical games,
namely, in Brownian or molecular motors. The name ``molecular
motor" comes from biophysics where several forms of energy
conversion within the cell are studied. For example, there
are``pumps" capable of transporting ions from one side of a
biological membrane to the other against the ion's electrochemical
gradient \cite{alberts94,cooper97}; molecules travelling through
microtubules transporting diverse cellular material; or molecules
capable of ``pulling off" a filament and making up the muscular
tissue \cite{cooper97}. Each one of these systems consists of a
unique protein that changes its tri-dimensional configuration upon
hydrolysis of ATP molecules. The protein is different for each
system considered (kinesins in the transport through
micro-channels, myosines in the muscular tissue), but all of them
are motors capable of extracting the stored energy in the ATP
bonds and transform it into mechanical energy. All these
transformations are carried out at a scale where the thermal
fluctuations are unavoidable.

For all this, we believe that in order to understand the
energy-exchange processes at the cellular level a disposal of the
authentic thermodynamics of Brownian motors will be essential.
This thermodynamics will have to be, as we have seen, one of
irreversible processes wherein the thermal fluctuations play a
relevant role. It will, also, have to be different from the
so-called finite-time thermodynamics, since the latter studies
irreversible motors but of macroscopic character.

In recent years, the first steps have been taken towards building
up the theory of molecular motors. Nevertheless, despite some work
that shows general properties
\cite{parrondo98b,parrondo98a,parrondo96,leibler93,magnasco94}
most of the results so far refer to concrete models
\cite{astumian94,sokolov97} that are too simple to represent the
behaviour of molecular motors.

Finally, we would like to conclude by recalling that the
paradoxical games may have applications in other contexts. The
paradox shows that the result of alternating two stochastic
dynamics is far from being the ``sum" or combination of the
effects of each dynamic considered alone, and that could, in fact,
be completely unexpected. This result should raise the interest in
alternating dynamics in physical, biological and economics
systems.

\section*{Appendix 1 -- Markov Chain}

A Markov chain is a set of $N$ states and we consider a particle
that jumps in from one state to another in a probabilistic way and
in discrete time steps. The Markov chain is defined by the matrix

\begin{equation}
\label{} \Pi=\{p_{i\rightarrow j}\}^N_{i,j=1}
\end{equation}

where $p(i\rightarrow j)$ is the particle transition probability
from state $i$ to the state $j$. Contrary to the usual matrix
notation we shall use $i$ as the column index and $j$ as the row
index. It can be readily checked that all the matrix elements are
real numbers between $0$ and $1$ and that the entries of every
column add up to $1$.

Let us call $\pi_i(t)$ the probability that the particle is in the
state $i$ in time $t$. The probability to find the particle in the
state $i$ at time $t+1$ is equal to the probability that the
particle jumps, in the step $t\rightarrow t+ 1$, from any other
state $j$ (including the case $i=j$), that is

\begin{equation}
\label{equation12}
\pi_{i}(t+1)=\sum^N_{j=1}\pi_{j}(t)p_{j\rightarrow i}.
\end{equation}

Hence the distribution of probability $\pi(t)=(\pi_1(t),\ldots
\pi_N(t))$ satisfies the following evolution equation

\begin{equation}
\label{equation13} \vec{\pi}(t+1)=\Pi\vec{\pi}(t).
\end{equation}

The distribution $\pi(t)$ tends, at large $t$, to a stationary
distribution, $\pi^{\mathrm{st}}_i$, that satisfies equation
(\ref{equation14})

\begin{equation}
\label{equation14}
\vec{\pi}^{\mathrm{st}}(t+1)=\Pi\vec{\pi}^{\mathrm{st}}(t)
\end{equation}

that is, $\pi^{\mathrm{st}}$ is the eigenvector of the matrix
$\Pi$ with eigenvalue $1$.

The stationary distribution tells us, in the limit of long times,
what are the probabilities of the particle being in the different
states $i=1,2,\ldots N$.

Game B consists of a Markov chain of three states, namely, $0$,
$1$ and $2$. The matrix $\Pi$ is

\begin{equation}
\label{equation15}
\Pi= \left(\begin{array}{ccc}
  0 & 1-p_2 & p_2 \\
  p_3 & 0 & 1-p_2 \\
  1-p_3 & p_2 & 0 \\
\end{array}\right).
\end{equation}

Notice that the elements of every column add up to $1$. The
stationary probability distribution is

\begin{equation}
\label{equation16}
\vec{\pi}^{\mathrm{st}}=\frac{1}{Z}\left(1,\frac{p_2p_3+1-p_2}{1-p_2+p^2_2},
\frac{p_2p_3+1-p_3}{1-p_2+p^2_2}\right)
\end{equation}

where $Z$ is a normalization constant. Finally, a similar analysis
holds for the random combination of games A and B upon
substituting $p_2$ for $\frac{(p_2+p_1)}{2}$ and $p_3$ for
$\frac{(p_3+p_1)}{2}$. From this stationary probability
distribution $\pi^{\mathrm{st}}_i$ it is easy to compute the
probability of winning in a particular run, as we have done in the
main text.

\section*{Appendix 2 -- The Metropolis algorithm}

Let us suppose we have a physical system that can be in any of the
states $i=1,2,\ldots N$ with energy $E_i$. What are the transition
probabilities $p(i\rightarrow j)$ that govern the probabilistic
evolution of the system when this is in contact with with a
thermal bath at temperature $T$?

The requirement these transition probabilities have to satisfy is
that they should eventually attain the Boltzmann distribution,
that is, its stationary distribution should be

\begin{equation}
\label{equation17}
\vec{\pi}^{\mathrm{st}}=\frac{1}{Z} e^{-\beta
E_i/kT}
\end{equation}

where $Z$ is a normalized constant and $\beta=\frac{1}{kT}$, with
$k$ the Boltzmann constant. Consequently, the transition matrix
$\Pi$ has to be such that it has the distribution
$\pi^{\mathrm{st}}$, given by (\ref{equation17}), as its
eigenvector with eigenvalue $1$.

A particular case is when the transition probabilities satisfy the
condition of detailed balance

\begin{equation}
\label{equation18} \pi^{\mathrm{st}}_{i}p_{i\rightarrow j}=
\pi^{\mathrm{st}}_{j}p_{j\rightarrow i}.
\end{equation}

An even more particular case is the Metropolis algorithm wherein
the transition probabilities are given by the following rule, for
$i\neq j$: or $i$ and $j$ are not connected, in which case
$p(i\rightarrow j)=p(j\rightarrow i)=0$, or they are connected, so
that

\begin{equation}
\label{equation19}
p_{i\rightarrow j}=\left\{
\begin{array}{cc}
\frac{1}{C_i}  \hspace{1.6cm}& \mathrm{if} \hspace{0.15cm} E_j\leq E_i,\\
\frac{1}{C_i}e^{-(E_j-E_i)\beta} & \mathrm{if} \hspace{0.15cm}
E_j\geq E_i,
\end{array}\right.
\end{equation}

where $C_i$ is chosen so that all the transition probabilities are
between $0$ and $1$ (it often is the coordination number, i.e. the
number of states connected to the state $i$) Finally, the
transition probability from $i$ to itself is

\begin{equation}
\label{equation20} p_{i\rightarrow i}=1-\sum_{j\neq
i}p_{i\rightarrow j}
\end{equation}

where the sum is carried over all the states $j$ connected to $i$.
It can readily be checked that the transition probabilities so
defined satisfy the condition of detailed balance.

We shall use the Metropolis algorithm in this article to model a
particle in contact with a thermal bath. Notice also that in
Equation (\ref{equation19}) there is a non-zero probability that
the particle increases its energy. These transitions are the
result of thermal fluctuations where the thermal bath gives energy
to the particle.

\end{document}